\documentclass[12pt]{article}
\usepackage{amsfonts,amsmath,amssymb,amscd}
\newtheorem{definition}{Definition}[section]

\newtheorem{theorem}[definition]{Theorem}

\newtheorem{lemma}[definition]{Lemma}

\def\cA{{\cal A}}          \def\cB{{\cal B}}          
                    
          \def\cH{{\cal H}}          \def\cI{{\cal I}}
                    \def\cL{{\cal L}}
                    
\def\cP{{\cal P}}          \def\cQ{{\cal Q}}          
\def\cS{{\cal S}}          \def\cT{{\cal T}}

\def\fS{{\mathfrak S}}

\newcommand{\ie}{{\it i.e.}\ }
\newcommand{\CC}{{\mathbb C}}

\newcommand{\ZZ}{{\mathbb Z}}
\newcommand{\RR}{{\mathbb R}}

\newcommand{\finproof}{{\hfill \rule{5pt}{5pt}}}

\def\QTHA{Quasitriangular Hopf Algebra (QTHA)\def\QTHA{QTHA}}
\def\QTQHA{Quasitriangular Quasi-Hopf Algebra (QTQHA)\def\QTQHA{QTQHA}}
\begin{document}
\pagestyle{empty}
\begin{center}

{\Large \textsf{Integrable $N$-particle Hamiltonians with
\\Yangian or Reflection Algebra Symmetry}}
\vspace{10mm}

{\large V.~Caudrelier and N.~Cramp{\'e}}

\vspace{10mm}

\emph{Laboratoire d'Annecy-le-Vieux de Physique Th{\'e}orique}\\
\emph{LAPTH, CNRS, UMR 5108, Universit{\'e} de Savoie}\\
\emph{B.P. 110, F-74941 Annecy-le-Vieux Cedex, France}\\

\vspace{7mm}
\end{center}
\vfill \vfill
\begin{abstract}
We use the Dunkl operator approach to construct one dimensional
integrable models describing $N$ particles with internal degrees
of freedom. These models are described by a general Hamiltonian
belonging to the center of the Yangian or the reflection algebra,
which ensures that they admit the corresponding symmetry. In
particular, the open problem of the symmetry is answered for the
$B_N$-type Sutherland model with spin and for a generalized
$B_N$-type nonlinear Schr{\"o}dinger Hamiltonian.
\end{abstract}
\vfill MSC number: 70H06, 81R12, 81R50 \vfill
\rightline{LAPTH-1002/03}
\rightline{math-ph/0310028}
\baselineskip=16pt
\newpage
\pagestyle{plain} \setcounter{page}{1}

\section*{Introduction}
\quad~The introduction of internal degrees of freedom in an
increasing number of one dimensional quantum integrable systems
has proved to be fruitful in various physical and mathematical
investigations. This is well illustrated in the study of
symmetries. In particular, the Yangian symmetry was exhibited in
the $A_N$ Sutherland model with spin \cite{BGHP}, the $A_N$
confined Calogero model with spin \cite{H} or the quantum
nonlinear Schr\"odinger (NLS) equation \cite{MW,MRSZ}. This in
turn allows to find the spectrum and degeneracies.

The main idea of this article is to generalize the Dunkl operator
approach of \cite{BGHP} in order to construct a general N-body
Hamiltonian which possesses the reflection algebra \cite{Skl} as
symmetry algebra. A direct consequence is the integrability of the
system described by this general Hamiltonian. Taking a particular
case, we answer the question of the symmetry of the $B_N$
Sutherland model with spin. In the same way, we exhibit the
symmetry of a generalized $B_N$-type NLS Hamiltonian. With the
same procedure, we also construct a general N-body integrable
Hamiltonian with Yangian symmetry from which we recover the known
cases of NLS and $A_N$ Sutherland model with spin.

After recalling some known mathematical background needed in the
construction of the central elements of the Yangian \cite{Dri85}
of $gl(n)$, $Y(n)$, and of the reflection algebra, $\cB(n)$, in
section \ref{central}, we give a realization of these algebras in
terms of transfer matrices and generators of the \textit{extended}
degenerate affine Hecke algebra, $\cA(N)$. Next, we prove the main
theorems of section \ref{real} which provide another realization
for each algebra $\cB(n)$ and $Y(n)$ in terms of a projector
specifying the physical properties of the wave functions occurring
when we represent our setup in section \ref{ham}. We identify a
central element used in section \ref{ham} (resp. section
\ref{hamY}) to construct the general one dimensional $N$-particle
Hamiltonian for which we prove integrability and reflection
algebra (resp. Yangian) symmetry. This is done by representing
$\cA(N)$ in terms of operators (in particular Dunkl operators)
acting on the space of wave functions. Then, we particularize the
former general Hamiltonian and conclude on the symmetry of
generalizations of NLS and Sutherland models.

\section{Central elements of $Y(n)$ and $\cB(n)$}\label{central}
\setcounter{equation}{0} We deal with the multiple tensor products
$\big(End(\CC^n)\big)^{\otimes m}$ where $m\in \ZZ_{\ge 0}$ will
be the number of copies necessary for the equations to make sense.
For $A\in End(\CC^n)$ and $k\in\{1,\dots,m\}$, we define $A_k$ by
\begin{eqnarray}
A_k=1^{\otimes k-1}\otimes A \otimes  1^{\otimes m-k}
~~\in~~\big(End(\CC^n)\big)^{\otimes m}\;.
\end{eqnarray}

\subsection{Yangian $Y(n)$}\label{def_Y}
\setcounter{equation}{0}

The Yangian of $gl_n$ \cite{Dri85}, $Y(n)$,
is the complex associative algebra, generated by the unit and the
elements $\{t_{ij}^{(k)}\;|\;1\le i,j \le n;\,k\in\ZZ_{>0}\}$
gathered in the formal series
\begin{eqnarray}
t_{ij}(u)
  =\delta_{ij}+\lambda\;\sum_{k \in \ZZ_{> 0}}
  t_{ij}^{(k)} \, u^{-k}\in Y(n)[[u^{-1}]]
\end{eqnarray}
subject to the defining relations
\begin{eqnarray}
\label{RTT_relcom} &&(u-v)\left[{t}_{ij}(u),{t}_{kl}(v)\right]
=\lambda\big({t}_{kj}(u)\,{t}_{il}(v)-{t}_{kj}(v)\,{t}_{il}(u)\big),
\end{eqnarray}
where $\lambda$ is the parameter of deformation of the Yangian.
Let $E_{ij}$ be the elementary matrix with entry 1 in row $i$ and
column $j$ and zero elsewhere and $T(u)$ be defined by
\begin{eqnarray}
T(u)=\sum_{i,j=1}^{n}{t}_{ij}(u)\otimes E_{ij}\in
Y(n)[[u^{-1}]]\otimes End(\CC^n).
\end{eqnarray}
Then the relations (\ref{RTT_relcom}) are equivalent to the $RTT$
relation \cite{FRT}
\begin{eqnarray}
\label{RTT} && R_{12}(u-v)T_1(u)T_{2}(v)
=T_{2}(v)T_1(u)R_{12}(u-v)
\end{eqnarray}
where
\begin{eqnarray}
R_{12}(u)=1\otimes 1-\lambda\frac{P_{12}}{u}, \quad
P_{12}=\sum_{i,j=1}^{n}E_{ij}\otimes E_{ji}\in End(\CC^n)\otimes
End(\CC^n)
\end{eqnarray}
$P_{12}$ is the
permutation operator
i.e. $P_{12}v\otimes w=w \otimes v$, with $v,w\in \CC^n$.\\
This R-matrix, called the Yang matrix, satisfies the following
properties
\begin{eqnarray}
\label{YBE} &&R_{12}(u-v)R_{13}(u)R_{23}(v)
=R_{23}(v)R_{13}(u)R_{12}(u-v)\quad\mbox{(Yang-Baxter
equation)}\\
\label{unitarity}
&&R_{12}(u)R_{12}(-u)=\frac{u^2-\lambda^2}{u^2}1 \otimes 1
\quad\mbox{(unitarity relation)}.
\end{eqnarray}
Let $A_m$ be the antisymmetrizer operator in $(\CC^n)^{\otimes m}$
i.e.
\begin{equation}
A_m(e_{i_1}\otimes\cdots\otimes e_{i_m})=
\sum_{\sigma\in\mathfrak{S}_m}
sgn(\sigma)~e_{i_{\sigma(1)}}\otimes\cdots\otimes
e_{i_{\sigma(m)}}
\end{equation}
where $\{e_i|1\le i \le n\}$ is the canonical basis of $\CC^n$ and
$1\le i_1,\dots, i_m \le n$. One can show (see e.g. \cite{MNO})
that the following identities hold
\begin{eqnarray}
\label{comAT} A_m\;T_1(u)\cdots
T_m(u-m\lambda+\lambda)&=&T_m(u-m\lambda+\lambda)\cdots
T_1(u)\;A_m
\end{eqnarray}
For $m=n$, $A_n$ becomes a one-dimensional operator in
$(\CC^n)^{\otimes n}$ and the element (\ref{comAT}) is then equal
to $A_n$ times a scalar series with coefficients in $Y(n)$ called
the quantum determinant. This reads
\begin{equation}
\label{qdet} A_n\;qdet\;T(u)=A_n\;T_1(u)\cdots
T_n(u-n\lambda+\lambda).
\end{equation}
A well-known result (see e.g. \cite{Molev2001}) is that the
coefficients of $qdet\,T(u)$ generate the center of $Y(n)$.

\subsection{Reflection algebra $\cB(n)$}
Let $Q \in End(\CC^n)$ be an operator such that $Q^2=1$. Let us
introduce $\widetilde{\cB}(n)$ the complex associative algebra
generated by the unit and the elements
$\{\tilde{s}_{ij}^{(k)}\;|\;1\le i,j \le n;\,k\in\ZZ_{\ge0}\}$
gathered in the formal series
\begin{eqnarray}
\tilde{s}_{ij}(u)
  =\sum_{k \in \ZZ_{\ge 0}}
  \tilde{s}_{ij}^{(k)} \, u^{-k}\in \widetilde{\cB}(n)[[u^{-1}]].
\end{eqnarray}
The defining relations are given by the reflection equation
\cite{Skl, BPS}
\begin{eqnarray}
\label{eqrefle}
R_{12}(u-v)\;\widetilde{S}_1(u)\;Q_1\;R_{12}(u+v)\;Q_1\;\widetilde{S}_{2}(v)
=\widetilde{S}_{2}(v)\;Q_1\;R_{12}(u+v)\;Q_1\;\widetilde{S}_1(u)\;R_{12}(u-v)
\end{eqnarray}
where
\begin{eqnarray}
\widetilde{S}(u)=\sum_{i,j=1}^{n}\tilde{s}_{ij}(u)\otimes
E_{ij}\in \widetilde{\cB}(n)[[u^{-1}]]\otimes End(\CC^n).
\end{eqnarray}
There exists a connection between $Y(n)$ and $\widetilde{\cB}(n)$.
\begin{theorem}{\bf\cite{Skl}}
\label{embedding} Let
$$B(u)=\sum_{k\ge 0}\frac{B^{(k)}}{u^k} ~\in End(\CC^n)\;[[u^{-1}]]$$
satisfy the relation (\ref{eqrefle}). Then, the map
\begin{eqnarray}
\nonumber
\phi:\widetilde{\cB}(n)&\longmapsto& Y(n)\\
\widetilde{S}(u)&\longrightarrow& S(u)\equiv
T(u)\,B(u)\,Q\,T^{-1}(-u)\,Q
\end{eqnarray}
defines an algebra homomorphism.
\end{theorem}
In this article, we consider the reflection algebra $\cB(n)$,
subalgebra of $Y(n)$, defined as the image of $\widetilde{\cB}(n)$ by $\phi$.\\
By the same procedure as in \cite{MR}, one can define the Sklyanin
determinant
\begin{equation}
\label{def_sdet} A_n sdet\,S(u)=A_n \prod_{1\le k\le
n-1}^{\longrightarrow} \Big(S_k(u+\lambda-k\lambda)
R_{k,k+1}(2u+\lambda(1-2k))\cdots R_{k,n}(2u+\lambda(2-k-n)) \Big)
S_n(u+\lambda-n\lambda)
\end{equation}
where the product is ordered i.e. $\displaystyle\prod_{1\le k\le
n-1}^{\longrightarrow} X_k=X_1 \cdots X_{n-1}$. Following
\cite{MR}, one can express the Sklyanin determinant in terms of
the quantum determinant
\begin{equation}
\label{center} sdet\,S(u)=\theta(u)\;qdet\,T(u)\;
\big(qdet\,T(-u+n\lambda-\lambda)\big)^{-1}
\end{equation}
where~~$\theta(u)=sdet\,B(u) ~\in \CC\,[[u^{-1}]]$.\\
{}From theorem \ref{embedding} and relation (\ref{center}), one
deduces that the coefficients of the Sklyanin determinant belong
to the center of $\cB(n)$, which will be fundamental in
establishing the reflection symmetry.

\section{Realizations of $Y(n)$ and $\cB(n)$}\label{real}
\setcounter{equation}{0} This section is the first step toward our
goal. By realizing the above algebras, we will identify what will
be interpreted as Hamiltonians in the next sections.

\subsection{Extended degenerate affine Hecke algebra}
Let $N\in \ZZ_{\ge 2}$. The extended degenerate affine Hecke
algebra, $\cA(N)$, is the complex associative algebra generated by
the unit and three sets of elements denoted $\{d_i\;|\;1\le i \le
N\}$, $\{\cP_{i,i+1}\;|\;1\le i \le N-1\}$ and $\{\cQ_{i}\;|\;1\le
i \le N\}$ subject to the defining relations
\begin{eqnarray}
\label{defP}
&&\cP_{i,i+1}\cP_{i+1,i+2}\cP_{i,i+1}=\cP_{i+1,i+2}\cP_{i,i+1}\cP_{i+1,i+2}\\
&&\cP_{i,i+1}^2=1\\
\label{echangePd} &&\cP_{i,i+1}\;d_k=
\begin{cases}
d_k\;\cP_{i,i+1} &  k\neq i,i+1\\
d_{i+1}\;\cP_{i,i+1}+\beta & k=i\\
d_{i}\;\cP_{i,i+1}-\beta & k=i+1
\end{cases}\\
\label{commutedd} &&[d_i,d_j]=0\\
&&\cQ_{i}^2=1\\
&&\cQ_{i}\cQ_{j}=\cQ_{j}\cQ_{i}\\
&&\cQ_{i}\;\cP_{k,k+1}=
\begin{cases}
\cP_{k,k+1}\;\cQ_{i} &  i\neq k,k+1\\
\cP_{k,k+1}\;\cQ_{k+1} & i=k\\
\cP_{k,k+1}\;\cQ_{k} & i=k+1
\end{cases}\\
\label{echangeQd}
 &&\cQ_{i}\;d_k=
\begin{cases}
d_k\;\cQ_{i} &  k<i\\
-d_{i}\;\cQ_{i}+\beta\;\sum_{j=i+1}^N \cP_{ij}\big(\cQ_{i}+\cQ_{j}\big)+b & k=i\\
d_{k}\;\cQ_{i}+\beta\; \cP_{ik}\big(\cQ_{i}-\cQ_{k}\big) & k>i
\end{cases}
\end{eqnarray}
where
\begin{eqnarray}
&&\beta\in \CC\;,~~b\in \CC \quad\mbox{and}\\
&&\cP_{ij}=\cP_{i,i+1}\;\cP_{i+1,i+2}\cdots\cP_{j-2,j-1}\;\cP_{j-1,j}\;\cP_{j-2,j-1}
\cdots\cP_{i+1,i+2}\;\cP_{i,i+1}
\end{eqnarray}
The commutation relations (\ref{defP})-(\ref{echangeQd}) were
obtained in \cite{F02} for a particular representation but here we
set
them as abstract algebraic relations.\\
Let us note that the subalgebra of $\cA(N)$ generated by
$\{d_i|i=1,\dots,N\}$ and $\{\cP_{i,i+1}|i=1,\dots,N-1\}$
satisfying relations (\ref{defP})-(\ref{commutedd}) is the
degenerate affine Hecke algebra denoted $\widetilde{\cA}(N)$ first
introduced in \cite{Dri}.

\subsection{Transfer matrix}
In order to realize $Y(n)$ and $\cB(n)$ in terms of the elements of
$\cA(N)$, we suppose that the latter commute with $P$ and $Q$.
A realization of $Y(n)$ is given by the transfer matrix
\cite{BGHP}
\begin{equation}
\label{real_Y} \cT_0(u)= \cL_{01}(u) \cdots~ \cL_{0N}(u) \in
End(\CC^n)\otimes End(\CC^n)^{\otimes N}
\end{equation}
where
\begin{equation}
\cL_{0i}(u)= \frac{u+d_i}{u+d_i-\lambda} R_{0i}(u+d_i)=
\frac{u+d_i}{u+d_i-\lambda}
\left(1-\frac{\lambda\;P_{0i}}{u+d_i}\right).
\end{equation}
The first space denoted $0$ in (\ref{real_Y}) is called the
auxiliary space. The other ones, denoted $1,\dots,N$ and not
displayed explicitly in $\cT_0(u)$ for brevity, are called the
quantum spaces.

In the realization (\ref{real_Y}) of $Y(n)$, the quantum
determinant takes the following particular form
\begin{equation}
\label{form_qdet_serie}
qdet\,\cT(u)=\prod_{j=1}^N\frac{u+d_j}{u+d_j-n\lambda+\lambda}
\end{equation}
This realization allows us to obtain a realization of $\cB(n)$
thanks to theorem \ref{embedding} and relation (\ref{unitarity})
\begin{eqnarray}
\label{real_S}
\cS_0(u)&=&\cT_0(u)\,B_0(u)\,Q_0\,\cT_0^{-1}(-u)\,Q_0\\
&=&\frac{u+d_1}{u+d_1-\lambda}
\left(1-\frac{\lambda\;P_{01}}{u+d_1}\right)\cdots~
\frac{u+d_N}{u+d_N-\lambda}
\left(1-\frac{\lambda\;P_{0N}}{u+d_N}\right)
\,B_0(u)\,Q_0\nonumber\\
&&\hspace{1cm}\times~\frac{u-d_N}{u-d_N-\lambda}
\left(1-\frac{\lambda\;P_{0N}}{u-d_N}\right)\cdots~
\frac{u-d_1}{u-d_1-\lambda}
\left(1-\frac{\lambda\;P_{01}}{u-d_1}\right) Q_0\;
\end{eqnarray}
and one can compute
\begin{eqnarray}
\label{sdet} sdet\,\cS(u)&=&\!\!\theta(u)\prod_{j=1}^N
\frac{(u+d_j)(-u+d_j)}{(u+d_j-n\lambda+\lambda)(-u+d_j+n\lambda-\lambda)}
\\
\label{dev_sdet} &=&\theta_0
+\frac{1}{u}\Big(\theta_1+2(n\lambda-\lambda)N\theta_0\Big)
+\frac{1}{u^2}\Big(\theta_2+2(n\lambda-\lambda)N\theta_1
                   +(n\lambda-\lambda)^2N(2N+1)\theta_0\Big)
\nonumber\\
&&+\frac{1}{u^3}\Big(\theta_3+2(n\lambda-\lambda)N\theta_2
                   +(n\lambda-\lambda)^2N(2N+1)\theta_1\nonumber\\
            &&\hspace{1cm}+(n\lambda-\lambda)^3\frac{2N(N+1)(2N+1)}{3}\theta_0
                   +2\theta_0\;\cH\;\Big)+O\left(\frac{1}{u^4}\right)
\end{eqnarray}
where
\begin{equation}
\label{hamiltonien_di} \cH=\sum_{i=1}^N d_i^2
\end{equation}
and the coefficients $\theta_j~(j=0,1,2,3)$ are given by the
expansion
\begin{equation}
\label{theta}
\theta(u)=sdet\,B(u)=\theta_0+\frac{\theta_1}{u}+\frac{\theta_2}{u^2}
+\frac{\theta_3}{u^3}+O\left(\frac{1}{u^4}\right)\;.
\end{equation}
As announced earlier, we identified a central element $\cH$ whose
interpretation as Hamiltonian will become explicit in sections
\ref{ham} and \ref{hamY}.

\subsection{Projectors}\label{projectors}

We now turn to a crucial point in our construction. Let us define
two operators
\begin{eqnarray}
\label{L1} \Lambda^{(1)}&=&\frac{1}{N!}\prod_{j=2}^{N}\big(1+\tau'
P_{1j}\cP_{1j}
+\cdots+\tau' P_{j-1,j}\cP_{j-1,j}\big)\\
\label{L2}
\Lambda^{(2)}&=&\frac{1}{2^N}\prod_{j=1}^{N}\big(1+\tau''
Q_j\cQ_j\big)
\end{eqnarray}
where $\tau', \tau'' = \pm 1$. We define
$\Lambda=\Lambda^{(1)}\Lambda^{(2)}=\Lambda^{(2)}\Lambda^{(1)}$.
One can check that the operators $\Lambda^{(1)}$, $\Lambda^{(2)}$
and $\Lambda$ are projectors. Let us remark that the products in
relations (\ref{L1}) and (\ref{L2}) are not necessarily ordered
since the factors in each product commute with one another.
\begin{lemma}
\label{P_K} For $1\le i< j \le N$ and $1\le l \le N$, one has
\begin{eqnarray}
\label{P_K1}
\left(1-\tau' P_{ij}\cP_{ij}\right)\Lambda^{(1)}&=&0\\
\label{P_K2} \left(1-\tau''
Q_{l}\cQ_{l}\right)\Lambda^{(2)}&=&0\;.
\end{eqnarray}
\end{lemma}
\textbf{Proof:} Let $\sigma\in\fS_N$. An equivalent definition of
$\Lambda^{(1)}$ is
$$\Lambda^{(1)}=\frac{1}{N!}\prod_{k=2}^{N}
\big(1+\tau' P_{\sigma(1)\sigma(k)}\cP_{\sigma(1)\sigma(k)}
+\cdots+\tau'
P_{\sigma(k-1)\sigma(k)}\cP_{\sigma(k-1)\sigma(k)}\big).$$ For
$1\le i< j \le N$, let us choose $\sigma$ so that $\sigma(1)=i$
and $\sigma(2)=j$. Then, one gets
\begin{eqnarray}
&&\hspace{-2cm}(1-\tau' \cP_{ij}P_{ij})\Lambda^{(1)}= (1-\tau'
\cP_{ij}P_{ij})(1+\tau' \cP_{ij}P_{ij})
\nonumber\\
&&\hspace{2cm}\times\frac{1}{N!}\prod_{k=3}^{N} \big(1+\tau'
P_{\sigma(1)\sigma(k)}\cP_{\sigma(1)\sigma(k)} +\cdots+\tau'
P_{\sigma(k-1)\sigma(k)}\cP_{\sigma(k-1)\sigma(k)}\big)=0
\end{eqnarray}
which proves relation (\ref{P_K1}). Relation (\ref{P_K2}) is
straightforward.
\finproof\\
In the rest of this article, we take a particular form for $B(u)$
\begin{equation}
\label{formB} B(u)=1+b'\frac{Q}{u}\quad (b'\in\CC)\;.
\end{equation}
In this case, the constant coefficient $\theta_0$ in (\ref{theta})
is $1$. Let us now state the main theorem of this section.
\begin{theorem}
\label{projec2} If $\beta=\tau'\lambda$ and $b=-2\tau''b'$, then
$\cS(u)\Lambda$ is a realization of $\cB(n)$ i.e. one gets
\begin{eqnarray}
\label{refle1} \hspace{-0.4cm}
R_{00'}(u-v)\;\cS_0(u)\Lambda\;Q_0\;
R_{00'}(u+v)\;Q_0\;\cS_{0'}(v)\Lambda
=\cS_{0'}(v)\Lambda\;Q_0\;R_{00'}(u+v)\;
Q_0\;\cS_0(u)\Lambda\;R_{00'}(u-v)
\end{eqnarray}
The Sklyanin determinant can be computed thanks to the following
formula
\begin{eqnarray}
\label{Lsdet}
sdet\,\big(\cS(u)\Lambda\big)&=&\big(sdet\,\cS(u)\big)\Lambda
\end{eqnarray}
\end{theorem}
\textbf{Proof:} Noting that $\Lambda$ commutes with $R_{00'}$ and
$Q_0$, the validity of relation (\ref{refle1}) is implied by
\begin{eqnarray}
\label{LambdaS} (\Lambda-1)\cS_0(u)\Lambda=0\, .
\end{eqnarray}
This in turn holds if
\begin{equation}
\begin{cases}
(\cP_{i,i+1}-\tau' P_{i,i+1})~\cS_0(u)\Lambda^{(1)}=0~,~~~i=1,\ldots,N-1\\
(\cQ_{N}-\tau'' Q_{N})~\cS_0(u)\Lambda^{(2)}=0
\end{cases}
\end{equation}
Now a direct computation using the exchange relations of $\cA(N)$
and the conditions on $\beta$ and $b$ allows one to find $\cS'$
and $\cS''$ such that
\begin{equation}
\begin{cases}
(\cP_{i,i+1}-\tau' P_{i,i+1})~\cS_0(u)=\cS'_0(u)~(\cP_{i,i+1}-\tau' P_{i,i+1})\\
(\cQ_{N}-\tau'' Q_{N})~\cS_0(u)=\cS''_0(u)~(\cQ_{N}-\tau'' Q_{N})
\end{cases}
\end{equation}
which finishes the proof of (\ref{LambdaS}) invoking lemma
\ref{P_K}. Relation (\ref{Lsdet}) is proven using the definition
(\ref{def_sdet}) of the Sklyanin determinant and relation (\ref{LambdaS}). \finproof\vspace{3mm}\\
\textit{Remark:} One can verify that the validity of
(\ref{refle1}) actually imposes the explicit form (\ref{formB}) of
$B(u)$ up to a normalization and the above constraints on
$\lambda$ and $b'$.

In a similar way, one can prove the following theorem. The latter
encompasses the analog result in \cite{BGHP}. Indeed, one recovers
the situation of \cite{BGHP} by specifying a particular
representation of the generators of $\cA(N)$.
\begin{theorem}
\label{projec1} If $\beta=\tau'\lambda$, then
$\cT(u)\Lambda^{(1)}$ is a realization of $Y(n)$ i.e. one gets
\begin{eqnarray}
\label{YBE_1}
 R_{00'}(u-v)~\cT_0(u)\Lambda^{(1)}~\cT_{0'}(v)\Lambda^{(1)}
=\cT_{0'}(v)\Lambda^{(1)}~\cT_0(u)\Lambda^{(1)}~R_{00'}(u-v).
\end{eqnarray}
The quantum determinant of $\cT(u)\Lambda^{(1)}$ can be computed
thanks to the following formula
\begin{eqnarray}
\label{qdet1} qdet\,\big(\cT(u)\Lambda^{(1)}\big)
&=&\big(qdet\,\cT(u)\big)\Lambda^{(1)}
\end{eqnarray}
\end{theorem}
\textbf{Proof:} The proof is similar to that of theorem \ref{projec2}
. \finproof\vspace{3mm}\\

\section{Hamiltonians with $\cB(n)$ symmetry}
\label{ham} \setcounter{equation}{0}

In this section and the next one, we present the physical
application of the above mathematical setting. We will work in the
first quantized picture with $N$ indistinguishable particles. Let
$\{q_i|1\le i \le N\}$ be the coordinates and $\{s_i|1\le i \le
N\}$ the internal degrees of freedom (or spins) of the particles.
Any $s_i$ takes values in $\Sigma=\{-\frac{n-1}{2},-\frac{n-3}{2},
\dots,\frac{n-3}{2},\frac{n-1}{2}\}$. Then, the wave function of
the system is denoted $\phi(q_1,\cdots,q_N|s_1,\cdots,s_N)$.

\subsection{Representation of $\cA(N)$ and associated
Hamiltonians}

We represent $P$, $Q$ and the generators of $\cA(N)$ as operators
on the space $\mathfrak{L}$ of wave functions. This reads, for
$1\le i < j \le N$ and $\phi\in\mathfrak{L}$,
\begin{eqnarray}
\label{rep_cP}
&&\cP_{ij}~\phi(q_1,\cdots,q_i,\cdots,q_j,\cdots,q_N|s_1,\cdots,s_N)
=\phi(q_1,\cdots,q_j,\cdots,q_i,\cdots,q_N|s_1,\cdots,s_N)\\
\label{rep_P}
&&P_{ij}~\phi(q_1,\cdots,q_N|s_1,\cdots,s_i,\cdots,s_j,\cdots,s_N)
=\phi(q_1,\cdots,q_N|s_1,\cdots,s_j,\cdots,s_i,\cdots,s_N)
\end{eqnarray}
\ie $\cP_{ij}$ (resp. $P_{ij}$) is the permutation operator acting
on positions (resp. spins) of the $i^{th}$ and $j^{th}$ particles.
And for $1 \le i \le N$, we define
\begin{eqnarray}
\label{rep_cQ}
&&\cQ_{i}~\phi(q_1,\cdots,q_i,\cdots,q_N|s_1,\cdots,s_N)
=\phi(q_1,\cdots,\alpha(q_i),\cdots,q_N|s_1,\cdots,s_N)\\
\label{rep_Q}
&&Q_{i}~\phi(q_1,\cdots,q_N|s_1,\cdots,s_i,\cdots,s_N)
=\phi(q_1,\cdots,q_N|s_1,\cdots,s_i^*,\cdots,s_N)\;.
\end{eqnarray}
where $\alpha$ is a function defining the action of $\cQ_i$ on the
position of the $i^{th}$ particle and $*$ represents the action of
$Q_i$ on its spin. Since $\cQ_i^2=1$ and $Q_i^2=1$ , one gets
$\alpha(\alpha(q_i))=q_i$ and $(s_i^*)^*=s_i$. Now, we choose
$d_l$ to be a Dunkl operator \cite{D} defined as follows, for
$1\le l \le N$,
\begin{equation}
\label{dunklB} d_l=a(q_l)\frac{\partial}{\partial q_l}
+\sum_{k=1}^{l-1}v(q_l,q_k)\cP_{kl}-\sum_{k=l+1}^{N}v(q_k,q_l)\cP_{lk}
+\sum_{k=1,k\neq
l}^{N}\overline{v}(q_l,q_k)\overline{\cP}_{lk}+g(q_l)\cQ_l
\end{equation}
where $\overline{\cP}_{lk}=\cQ_l\,\cQ_k\,\cP_{lk}$. For the
product of Dunkl operators to be well-defined, $a$, $v$,
$\overline{v}$, $g$ must be $C^\infty$ functions.
\begin{theorem}\label{solution}
For $a\neq 0$ and $A(x)=\int^x\frac{dy}{a(y)}$ invertible, the
operators $\cP_{ij}$, $\cQ_{i}$ and $d_i$ as defined in
(\ref{rep_cP}), (\ref{rep_cQ}) and (\ref{dunklB}) realize $\cA(N)$
if and only if
\begin{eqnarray}
&&\alpha(x)=A^{-1}(-A(x))\\
&&v(x,y)=\frac{\beta}{e^{-2\gamma\big(A(x)-A(y)\big)}-1}~, \quad
\gamma\in\CC\\
&&\overline{v}(x,y)=\frac{\beta}{1-e^{2\gamma\big(A(x)+A(y)\big)}}\\
&&g(x)=\frac{c-b\;e^{-2\gamma A(x)}}{2sinh(2\gamma A(x))}~, \quad
c\in\CC\;.
\end{eqnarray}
\end{theorem}
\textbf{Proof:} The constraints on the functions $\alpha$, $a$,
$v$, $\overline{v}$ and $g$ arise from (\ref{echangePd}),
(\ref{commutedd}) and (\ref{echangeQd}). Starting from
(\ref{commutedd}), the idea is to cancel the coefficients
appearing in front of independent operators such as $\cP_{ij}$ or
$\cP_{ik}\cP_{jk}$:
\begin{eqnarray}
&&a(x)\frac{\partial}{\partial
x}v(x,y)+a(y)\frac{\partial}{\partial
y}v(x,y)=0\\
&&-v(y,z)v(x,z)+v(x,y)v(y,z)+v(x,z)v(y,x)=0
\end{eqnarray}
whose solution is given by
$$v(x,y)=\frac{C}{e^{-2\gamma\big(A(x)-A(y)\big)}-1}~, \quad
C,\gamma\in\CC$$ and (\ref{echangePd}) imposes $C=\beta$. The form
of $\alpha$, $\overline{v}$ and $g$ are found in the same way.
Then, a global check ensures that all the remaining relations
are identically satisfied. \finproof\vspace{3mm}\\

The Dunkl operators realized as in (\ref{dunklB}) are independent
and from (\ref{commutedd}), (\ref{hamiltonien_di}), we have
\begin{equation}
\label{commuteHd} \left[\cH,d_i\right]=0\qquad\mbox{for}\quad
i=1,\dots,N\;,
\end{equation}
which ensures the integrability. Then, from
(\ref{hamiltonien_di}), we can compute
\begin{eqnarray}
\label{hamiltonienB}
\cH&=&\sum_{i=1}^{N}\left(a(q_i)^2\frac{\partial^2}{\partial
q_i^2} +a(q_i)\frac{\partial a(q_i)}{\partial
q_i}\frac{\partial}{\partial q_i} \right)\nonumber\\
& &+ \sum_{1\le i<j\le N}\left(\frac{\beta\gamma
\Big(\cP_{ij}-\frac{\beta}{2\gamma}\Big)} {sinh^2\left[\gamma\big(
A(q_i)- A(q_j)\big)\right]}+\frac{\beta\gamma
\Big(\overline{\cP}_{ij}-\frac{\beta}{2\gamma}\Big)}
{sinh^2\left[\gamma\big( A(q_i)+ A(q_j)\big)\right]} \right)\nonumber\\
& &+\sum_{i=1}^N\left(\frac{\gamma
(b+c)\left(\cQ_i-\frac{b+c}{4\gamma}\right)}{4\,sinh^2\left[\gamma
A(q_i)\right]}-\frac{\gamma
(b-c)\left(\cQ_i-\frac{b-c}{4\gamma}\right)}
{4\,cosh^2\left[\gamma A(q_i)\right]}\right)
\end{eqnarray}
Now the constructions of the previous sections get their physical
meaning. $\Lambda^{(1)}$ is the projector from $\mathfrak{L}$ onto
$\mathfrak{L}^{(1)}_{\tau'}$, the space of globally
$\tau'$-symmetric wave functions ($\tau'=1$ for symmetric and
$\tau'=-1$ for antisymmetric). $\Lambda^{(2)}$ is the projector
from $\mathfrak{L}$ onto $\mathfrak{L}^{(2)}_{\tau''}$, the space
of wave functions such that
\begin{equation}
\phi(q_1,\cdots,\alpha(q_i),\cdots,q_N|s_1,\cdots,s_i^*,\cdots,s_N)
=\tau''\phi(q_1,\cdots,q_i,\cdots,q_N|s_1,\cdots,s_i,\cdots,s_N)
\end{equation}
And then, $\Lambda$ is the projector from $\mathfrak{L}$ onto
$\mathfrak{L}_{\Lambda}=\mathfrak{L}^{(1)}_{\tau'}\cap\mathfrak{L}^{(2)}_{\tau''}$.
\begin{theorem}\label{effective}
Let $\overline{P}_{ij}=Q_iQ_jP_{ij}$ and $c'=-\frac{c\tau''}{2}$.
The effective Hamiltonian, $\cH_{\Lambda}$, restricted to
$\mathfrak{L}_{\Lambda}$, reads
\begin{eqnarray}
\label{Hlambda}
\cH_{\Lambda}&=&\sum_{i=1}^{N}\left(a(q_i)^2\frac{\partial^2}{\partial
q_i^2} +a(q_i)\frac{\partial a(q_i)}{\partial
q_i}\frac{\partial}{\partial q_i} \right)\nonumber\\
& &+ \sum_{1\le i<j\le N}\left(\frac{\gamma\lambda
\Big(P_{ij}-\frac{\lambda}{2\gamma}\Big)} {sinh^2\left[\gamma\big(
A(q_i)- A(q_j)\big)\right]}+\frac{\gamma\lambda
\Big(\overline{P}_{ij}-\frac{\lambda}{2\gamma}\Big)}
{sinh^2\left[\gamma\big( A(q_i)+ A(q_j)\big)\right]} \right)\nonumber\\
& &+\sum_{i=1}^N\left(-\,\frac{\gamma
(b'+c')\left(Q_i+\frac{b'+c'}{2\gamma}\right)}
{2\,sinh^2\left[\gamma A(q_i)\right]}
+\frac{\gamma(b'-c')\left(Q_i+\frac{b'-c'}{2\gamma}\right)}
{2\,cosh^2\left[\gamma A(q_i)\right]}\right)
\end{eqnarray}
and admits the reflection algebra as symmetry algebra. This
ensures in particular that it is integrable.
\end{theorem}
\textbf{Proof:} $\cH_{\Lambda}$ is actually $\cH\Lambda$ for
$\beta=\tau'\lambda$ and $b=-2\tau'' b'$. Indeed, the explicit
form above is obtained for the values of $\beta$ and $b$ just
specified and substituting $\cP$ and $\cQ$ for $P$ and $Q$ in
(\ref{hamiltonienB}) according to (\ref{P_K1})-(\ref{P_K2}). When
one restricts to $\mathfrak{L}_{\Lambda}$, $\Lambda$ is no longer
required on the right hand side of (\ref{Hlambda}). Then, relation
(\ref{dev_sdet}) and theorem \ref{projec2} imply that
$\cH_{\Lambda}$ admits the reflection algebra symmetry.\\
Integrability is proved upon expanding the Sklyanin determinant.
One can show that it can be written as
\begin{eqnarray}
\label{expansion}
sdet~(\cS(u)\Lambda)=\Lambda+\sum_{k=0}^{+\infty}\frac{1}{u^{k+1}}
\left[\lambda(n-1)\sum_{i=1}^N\left(1+(-1)^k\right)
d_i^k+G_k(d_1,\dots,d_N)\right]\Lambda
\end{eqnarray}
where $G_k$ is a $N$-variable polynomial of highest degree $k-1$.
We denote by $\cI_k$ the term between brackets in
(\ref{expansion}). Since the coefficients of the Sklyanin
determinant are central elements, one deduces that
\begin{equation}
\left[\cI_k\Lambda,\cI_l\Lambda\right]=0~~\text{and}~~\left[\cI_k\Lambda,\cH_{\Lambda}\right]=0,~~\forall
~k,l\in \ZZ_{\ge 0}
\end{equation}
and by paying attention to the terms of highest order in the
partial derivatives in $\cI_k\Lambda$, it is readily seen that
$\{\cI_{2k}\Lambda\}_{1\le k\le N}$ are independent, which proves
the integrability.\finproof

\subsection{Physical Hamiltonians and gauge fixing}

We still have to refine the form of the above Hamiltonian
$\cH_{\Lambda}$ so that its physical interpretation will be
easier. The aim is to recover the usual physical Hamiltonian in
units of $\hbar^2/2m$
\begin{equation}
\label{H_jauge} H=-\sum_{i=1}^{N}\frac{\partial^2}{\partial
z_i^2}~+~V(z_1,...,z_N)
\end{equation}
for some potential $V$. This can be achieved by performing a gauge
transformation $\mu(\mathbf{q})$ and a change of variables
$\mathbf{q}=\xi(\mathbf{z})$ with
$\mathbf{q}=(q_1,\dots,q_N),~\mathbf{z}=(z_1,\dots,z_N)$
\begin{equation}
H=\left.\mu(\mathbf{q})\; \cH_{\Lambda} \;
\frac{1}{\mu(\mathbf{q})}\right|_{\mathbf{q}=\xi(\mathbf{z})}\;.
\end{equation}
We note that this does not affect the results about the symmetry
and the integrability.\\
To get (\ref{H_jauge}) from $\cH_{\Lambda}$ given in
(\ref{Hlambda}), the suitable transformations are
\begin{eqnarray}
\label{transfo1}
\xi(\mathbf{z})&=&\left(A^{-1}(iz_1),\dots,A^{-1}(iz_N)\right)\\
\label{transfo2}
\mu(\mathbf{q})&=&\prod_{1\le i\le
N}\sqrt{a(q_i)}
\end{eqnarray}
\begin{theorem}
Under the transformations (\ref{transfo1})-(\ref{transfo2}), the
potential $V$ in (\ref{H_jauge}) splits into an external
potential, $U$, and a spin potential, $V_{spin}$,
\begin{equation}
V(\mathbf{z})=V_{spin}(\mathbf{z})~+~\sum_{k=1}^{N} U(z_k)\;.
\end{equation}
with
\begin{eqnarray}
\label{Vspin} V_{spin}(\mathbf{z})&=&-~\sum_{1\le i<j\le
N}\left(\frac{\gamma\lambda
\Big(P_{ij}-\frac{\lambda}{2\gamma}\Big)} {sin^2\left[\gamma\big(
z_i- z_j\big)\right]}+\frac{\gamma\lambda
\Big(\overline{P}_{ij}-\frac{\lambda}{2\gamma}\Big)}
{sin^2\left[\gamma\big( z_i+ z_j\big)\right]} \right)\nonumber\\
& &+~\sum_{i=1}^N\left(\frac{\gamma
(b'+c')\left(Q_i+\frac{b'+c'}{2\gamma}\right)}
{2\,sin^2\left(\gamma z_i\right)}
+\frac{\gamma(b'-c')\left(Q_i+\frac{b'-c'}{2\gamma}\right)}
{2\,cos^2\left(\gamma z_i\right)}\right)
\end{eqnarray}
and
\begin{equation}
\label{U}
U(x)=\frac{1}{4}a'\big(A^{-1}(ix)\big)^2
-\frac{1}{2}a\big(A^{-1}(ix)\big)a''\big(A^{-1}(ix)\big)
\end{equation}
where $a'(x)=\frac{d\,a(x)}{dx}\;.$
\end{theorem}
\textbf{Proof:} By direct computation\finproof\vspace{3mm}\\
To complete our discussion, we have to specify how the wave
function and the relations (\ref{rep_cP}), (\ref{rep_cQ})
transform under the change of variables (\ref{transfo1}). The wave
function $\phi'$ on which $H$ acts is given by
\begin{equation}
\phi'(z_1,\dots,z_N|s_1,\dots,s_N)=\phi(A^{-1}(iz_1),\dots,A^{-1}(iz_N)|s_1,\dots,s_N)
\end{equation}
It is then straightforward to see that the action of $\cP$ is
unchanged
$$\cP_{ij}~\phi'(z_1,\cdots,z_i,\cdots,z_j,\cdots,z_N|s_1,\cdots,s_N)
=\phi'(z_1,\cdots,z_j,\cdots,z_i,\cdots,z_N|s_1,\cdots,s_N)$$ and,
noting that $\alpha(A^{-1}(iz))=A^{-1}(-iz)$, the action of $\cQ$
reads
\begin{equation}
\label{actionQ}
\cQ_{i}~\phi'(z_1,\cdots,z_i,\cdots,z_N|s_1,\cdots,s_N)
=\phi'(z_1,\cdots,-z_i,\cdots,z_N|s_1,\cdots,s_N)
\end{equation}
\ie it is independent of $\alpha$ when we work with the variables
$z_i$. For wave functions in $\mathfrak{L}_{\Lambda}$, this implements
the Neumann (resp. Dirichlet) boundary condition for $\tau''=1$ (resp. $\tau''=-1$).\\
{}\\
We can give some comments on the form of the potentials. The term
$\gamma\lambda
\Big(P_{ij}-\frac{\lambda}{2\gamma}\Big)/\left(sin\left[\gamma\big(
z_i- z_j\big)\right]\right)^2$ expresses the usual two-body
interaction between the $i^{th}$ and $j^{th}$ particles and does
not break translation invariance as expected. The additional terms
can be better interpreted if one imagines a "mirror" sitting at
the origin $z=0$. Then, the term $\gamma\lambda
\Big(P_{ij}-\frac{\lambda}{2\gamma}\Big)/\left(sin\left[\gamma\big(
z_i+z_j\big)\right]\right)^2$ represents the two-body interaction
between the $i^{th}$ particle and the "mirror-image" of the
$j^{th}$ particle. And the remaining terms involving only $z_i$
accounts for the potential of the "impurity" at the origin. These
terms clearly violate translation invariance. Indeed, defining the
total momentum as usual
\begin{equation}
\label{impulsion}
\cI=-i\sum_{i=1}^N \frac{\partial}{\partial z_i}
\end{equation}
it is readily seen that
\begin{eqnarray}
\left[\cI,H\right]\neq 0\;.
\end{eqnarray}
We want to emphasize that this interpretation in terms of an
impurity sitting at the origin and of a "mirror-image" of the
system is possible thanks to (\ref{actionQ}), which is actually
related to the fact that the Hamiltonian $H$ is invariant under
the space reflections $z_i\to -z_i,~i=1,\ldots,N$.

\subsection{Examples}\label{known}

In all the above constructions, we have some freedom with the
function $a$ and the constants $\gamma$ and $c'$. In this section,
we use this freedom to exhibit particular Hamiltonians admitting
the reflection algebra as symmetry algebra.\\
We work with the Hamiltonian (\ref{H_jauge}) and from the previous
section, we know that we control the external potential $U$ thanks
to $a$ irrespective of $V_{spin}$. Thus, we suppose that the
function $a$ is constant so that the scalar external potential,
$U$, vanishes.

\subsubsection{$B_N$-type Nonlinear Schr{\"o}dinger Hamiltonian}

Let
\begin{equation}
\gamma=i\gamma'~,\quad\lambda=ig~,\quad b'=-ib_1,\quad
\mbox{where}~~\gamma',g,b_1\in\RR
\end{equation}
Taking the limit $\gamma'\rightarrow +\infty$ in (\ref{Vspin}) in
the sense of distributions, we get
\begin{equation}
H_{NLS}=-\sum_{k=1}^N \frac{\partial^2}{\partial
z_k^2}+2g\sum_{1\le k<l\le
N}\left[\delta(z_k-z_l)P_{kl}+\delta(z_k+z_l)\overline{P}_{kl}\right]
+2b_1\sum_{k=1}^N\delta(z_k)Q_k
\end{equation}
We know from the above results that this Hamiltonian admits the
reflection algebra symmetry and is integrable. Let us note that
when acting on $\mathfrak{L}_{\Lambda}$, we can drop the spin
operators $P_{ij},\overline{P}_{ij},Q_i$ in this particular case
due to the presence of the delta functions
\begin{equation}
H_{NLS}=-\sum_{k=1}^N \frac{\partial^2}{\partial
z_k^2}+2g\tau'\sum_{1\le k<l\le
N}\left[\delta(z_k-z_l)+\delta(z_k+z_l)\right]
+2b_1\tau''\sum_{k=1}^N\delta(z_k)
\end{equation}
This is the Hamiltonian of a system of $N$ bosonic ($\tau'=1$) or
fermionic ($\tau'=-1$) particles interacting through a delta
potential with coupling constant $g$ in the presence of a
delta-type impurity sitting at the origin.

\subsubsection{$B_N$ trigonometric/hyperbolic
Sutherland model with spin}

To recover the known integrable Hamiltonian of the $B_N$
trigonometric Sutherland model with spin \cite{F01_2}, we take
particular values of the constants present in
(\ref{H_jauge})-(\ref{Vspin})
\begin{eqnarray}
\gamma=1~,\quad\lambda=2g~,\quad b'+c'=-2b_1~~and\quad
b'-c'=-2b_2\quad\mbox{where}~~g,b_1,b_2\in\RR\;.
\end{eqnarray}
Thus, the Hamiltonian (\ref{H_jauge}) becomes
\begin{eqnarray}
\label{H_BtS} H_{BtS}=-\sum_{i=1}^{N} \frac{\partial^2}{\partial
z_i^2} &-&2g\sum_{1\le i<j\le N} \left(\frac{ \Big(P_{ij}-g\Big)}
{sin^2\big(z_i- z_j\big)} +\frac{\Big(\overline{P}_{ij}-g\Big)}
{sin^2\big( z_i+ z_j\big)}\right)\nonumber\\
&-&\sum_{i=1}^N \left(\frac{b_1\left(Q_i-b_1\right)}
{sin^2\left(z_i\right)} +\frac{b_2\left(Q_i-b_2\right)}
{cos^2\left(z_i\right)}\right)\;.
\end{eqnarray}
$g$ is the coupling constant and $b_1$, $b_2$ parametrize the
coupling with the impurity. From the general results of the
previous sections, we know that the reflection algebra is the
symmetry of the Hamiltonian
(\ref{H_BtS}).\\
The Hamiltonian of $B_N$ hyperbolic Sutherland model with spin
\cite{F02} is obtained by setting
\begin{eqnarray}
\gamma=i~,\quad\lambda=2ig~,\quad b'+c'=-2ib_1~~and\quad
b'-c'=2ib_2\quad \mbox{where}~~g,b_1,b_2\in\RR
\end{eqnarray}
and it takes the same form as (\ref{H_BtS}) but for the
trigonometric functions replaced by the corresponding hyperbolic
functions.

\section{Hamiltonians with $Y(n)$ symmetry}\label{hamY}
 \setcounter{equation}{0}
In this section, we take advantage of theorem \ref{projec1} and
just adapt all our machinery to exhibit a general integrable
Hamiltonian with Yangian symmetry which, once particularized,
reproduces already known systems such as nonlinear Schr{\"o}dinger
and $A_N$ Sutherland models with spin.

\subsection{Representation of $\widetilde{\cA}(N)$ and associated
Hamiltonians}

It is easy to see that $\sum_{i=1}^N d_i^2$ also appears in the
expansion of $qdet~\cT(u)$ in (\ref{form_qdet_serie}). As is
customary in the literature \cite{MW,Poly,BHV}, the starting point
is a representation of the degenerate affine Hecke algebra,
$\widetilde{\cA}(N)$. We keep (\ref{rep_cP}) and (\ref{rep_P}) and
take for the Dunkl operator
\begin{equation}
\label{dunklA}
d_l=a(q_l)\frac{\partial}{\partial q_l}
+\sum_{k=1}^{l-1}v(q_l,q_k)\cP_{kl}-\sum_{k=l+1}^{N}v(q_k,q_l)\cP_{lk}
\end{equation}
At this stage, we can reproduce along the same line the arguments
of section \ref{ham} to state the following theorems whose proofs
are similar to that of theorems \ref{solution}-\ref{effective} and
will not be given here.
\begin{theorem}
For $a\neq 0$ and $A(x)=\int^x\frac{dy}{a(y)}$ invertible, the
operators $\cP_{ij}$ and $d_i$ as defined in (\ref{rep_cP}) and
(\ref{dunklA}) realize $\widetilde{\cA}(N)$ if and only if
\begin{eqnarray}
v(x,y)=\frac{\beta}{e^{-2\gamma\big(A(x)-A(y)\big)}-1}~, \quad
\gamma\in\CC\;.
\end{eqnarray}
\end{theorem}
Again, we can construct the effective Hamiltonian
$\widetilde{\cH}_{\Lambda^{(1)}}$ whose properties are gathered in
\begin{theorem}
When restricted to $\mathfrak{L}^{(1)}_{\tau'}$, the effective
Hamiltonian
\begin{eqnarray}
\label{hamiltonienY}
\widetilde{\cH}_{\Lambda^{(1)}}=\sum_{i=1}^{N}\left(a(q_i)^2\frac{\partial^2}{\partial
q_i^2} +a(q_i)\frac{\partial a(q_i)}{\partial
q_i}\frac{\partial}{\partial q_i} \right)+\sum_{1\le i<j\le
N}\left(\frac{\gamma\lambda
\Big(P_{ij}-\frac{\lambda}{2\gamma}\Big)} {sinh^2\left[\gamma\big(
A(q_i)- A(q_j)\big)\right]}\right)
\end{eqnarray}
admits the Yangian symmetry and is integrable.
\end{theorem}
Now, performing the transformations
(\ref{transfo1})-(\ref{transfo2}) on
$\widetilde{\cH}_{\Lambda^{(1)}}$ we get the following physical
Hamiltonian
\begin{equation}
\label{hamY_jauge}
\tilde{H}=-\sum_{i=1}^{N}
\frac{\partial^2}{\partial
z_i^2}~+~\tilde{V}_{spin}(\mathbf{z})~+~\sum_{i=1}^{N} U(z_i)
\end{equation}
with $U$ given in (\ref{U}) and
\begin{equation}
\label{hamY_Vspin} \tilde{V}_{spin}(\mathbf{z})=-\sum_{1\le i<j\le
N}\frac{\gamma\lambda \Big(P_{ij}-\frac{\lambda}{2\gamma}\Big)}
{sin^2\left[\gamma\big( z_i- z_j\big)\right]}
\end{equation}
{\it Remark:} In the expansion of $qdet~\cT(u)$ in
(\ref{form_qdet_serie}), it is easy to see that there appears the
operator
\begin{equation}
\label{sum}
\sum_{i=1}^Nd_i=\sum_{i=1}^Na(q_i)\frac{\partial}{\partial q_i}\;.
\end{equation}
Assuming that $a$ is constant and performing the transformations
(\ref{transfo1})-(\ref{transfo2}), (\ref{sum}) becomes $\cI$ given
in (\ref{impulsion}). We then conclude that $\cI$ commutes with
our general Hamiltonian $\widetilde{H}$ so that the system is
translation invariant. In particular, this shows that the systems
we will consider in the next section with Yangian symmetry are
translation invariant as expected.

\subsection{Examples}

Using the freedom on $a$ and $\gamma$ in exactly the same fashion
as in section \ref{known}, we show that the Hamiltonian
(\ref{hamY_jauge}) generalizes known Hamiltonians for which the
Yangian symmetry and the integrability had already been proved:
\begin{itemize}
\item Nonlinear Schr{\"o}dinger Hamiltonian \cite{MW}
($\gamma=i\gamma'~,~~\lambda=ig~,~~\gamma',g\in\RR~,~~\gamma'\rightarrow
+\infty$)
\begin{equation}
\widetilde{H}_{NLS}=-\sum_{k=1}^N \frac{\partial^2}{\partial
z_k^2}+2g\tau'\sum_{1\le k<l\le N}\delta(z_k-z_l)
\end{equation}
\item $A_N$ trigonometric Sutherland model with spin
\cite{MP,F01_1} ($\gamma=1~,~~\lambda=2g~,~~g\in\RR$)
\begin{eqnarray}
\label{H_AtS} \widetilde{H}_{AtS}= -\sum_{i=1}^{N}
\frac{\partial^2}{\partial z_i^2} -2g\sum_{1\le i<j\le N}
\left(\frac{ \Big(P_{ij}-g\Big)} {sin^2\big(z_i- z_j\big)} \right)
\end{eqnarray}
\item $A_N$ hyperbolic Sutherland model with spin \cite{MP,F01_1}
($\gamma=i~,~~\lambda=2ig~,~~g\in\RR$)
\begin{eqnarray}
\label{H_AhS} \widetilde{H}_{AhS}= -\sum_{i=1}^{N}
\frac{\partial^2}{\partial z_i^2} -2g\sum_{1\le i<j\le N}
\left(\frac{ \Big(P_{ij}-g\Big)} {sinh^2\big(z_i- z_j\big)}
\right)
\end{eqnarray}
\end{itemize}

\section*{Conclusion and outlooks}

Starting from a representation of the {\it extended} degenerate
affine Hecke algebra in terms of operators acting on wave
functions, our main results are the construction of a general
$N$-particle Hamiltonian and the proof that it admits the
reflection algebra symmetry (theorems \ref{projec2} and
\ref{effective}). This ensures in particular its integrability.
The Yangian counterpart of this procedure gives back well-known
results.

The physical investigation of this Hamiltonian shows that it is
invariant under space reflections so that we considered wave
functions whose behaviour under the action of the operator
$\cQ_{i}$ is dictated by a parameter $\tau''=\pm 1$. This amounts
to encode a Neumann or Dirichlet boundary condition at $z=0$.
However, one sees that the "mirror-image" of the system on the
half-line is relevant and cannot be neglected if one wants to
maintain the usual nontrivial two-body interactions. Of course,
all this applies to the already known systems to which our general
Hamiltonian reduces in appropriate limits.

This brings us to the interesting issue of diagonalizing the
Hamiltonian $H$ using available results for reflection algebras.
This would provide the spectrum for apparently distinguished
models (such as $B_N$-type NLS or $B_N$ trigonometric/hyperbolic
Sutherland models), with boundary, unified by the Hamiltonian $H$.

\bigskip \textbf{Acknowledgements:} We warmly thank D.~Arnaudon,
L.~Frappat and E.~Ragoucy for helpful discussions and advice.

\end{document}